\input{aipcheck}

\documentclass[
    ,final            
  ]
  {aipproc}

\layoutstyle{6x9}

\usepackage{wrapfig}
\usepackage{graphicx}
\usepackage[centerlast]{caption}

\begin{document}

\title{An Exactly Solvable Supersymmetric Model of Semimagic
Nuclei}


\author{A. B. Balantekin}{
  address={Department of Physics, University of Wisconsin,
               Madison, Wisconsin 53706 USA}
}

\author{Nurta{\c{c}} G{\"{u}}ven}{
  address={Department of Applied Mathematics, Hali{\c{c}} University, Istanbul 34104 Turkey}
}

\author{Yama{\c{c}} Pehlivan}{
  address={Department of Applied Mathematics, Hali{\c{c}} University, Istanbul 34104 Turkey}
}

\begin{abstract}
A simple model of nucleons coupled to angular momentum zero
(s-pairs) occupying the valance shell of a semi-magic nuclei is
considered. The model has a separable, orbit dependent pairing
interaction which dominates over the kinetic term. It is shown
that such an interaction leads to an exactly solvable model whose
($0^+$) eigenstates and energies can be computed very easily with
the help of the algebraic Bethe ansatz method. It is also shown
that the model has a supersymmetry which connects the spectra of
some semimagic nuclei. The results obtained from this model for
the semimagic Ni isotopes from $^{58}$Ni to $^{68}$Ni are given.
In addition, a new and easier technique for calculating the energy
eigenvalues from the Bethe ansatz equations is also presented.
\end{abstract}

\maketitle


We consider the $s$-pairs of the same kind of nucleons (either
protons or neutrons) occupying the valance shell of a semi-magic
nuclei and interacting with an orbit dependent separable pairing
force as described by the Hamiltonian
\begin{equation}\label{1}
H=\sum_{j}\sum_m\varepsilon _{j}\> a_{j \> m}^\dagger a_{j \> m}
-|G|\sum_{jj^{\prime}}
c^*_{j}c_{j^{\prime}}S_{j}^{+}S_{j^{\prime}}^-.
\end{equation}
The orbit dependence of the pairing force is described by the
dimensionless constants $c_{j}$ while the overall strength of the
pairing term against the kinetic term is measured by the constant
$|G|$ which has the dimension of energy. In writing the
Hamiltonian (\ref{1}), we use the quasispin operators which are
given by \cite{Kerman1}
\begin{equation}\label{2}%
S_{j}^+=(S_{j}^{-})^\dagger=\sum_{m>0}(-1)^{j-m}a_{j \> m}^\dagger
a_{j \> -m}^\dagger \quad S_j^0=\frac{1}{2}\sum_{m>0}\left(a_{j \>
m}^\dagger a_{j \> m}+ a_{j \> -m}^\dagger a_{j \> -m} -1\right).
\end{equation}
Quasispin operators obey the angular momentum commutation
relations
\begin{equation}\label{3}
[S_j^+,S_{j^\prime}^-]=2\delta_{jj^\prime}S_j^0 \quad\quad
[S_j^0,S_{j^\prime}^\pm]=\pm\delta_{jj^\prime}S_j^\pm,
\end{equation}
i.e., one has an angular momentum algebra for each orbit $j$ such
that those angular momenta corresponding to different orbits
commute with one another. Note that Eq. (\ref{2}) implies $
-\Omega_j/2 \leq S_j^0 \leq \Omega_j/2$ where $\Omega_j=j+1/2$ is
the maximum number of $s$-pairs which can occupy the orbit $j$.
This tells us that the quasi-spin algebra corresponding to the
level $j$ is realized in the $\Omega_j/2$ representation.

Here, we are interested in the limit of the pairing Hamiltonian in
which all the single particle energy levels become degenerate. In
this \emph{degenerate limit}, the pairing Hamiltonian has two
interesting properties which are discussed in this talk: 1) It is
exactly solvable and 2) it is invariant under a supersymmetry
transformation which connects the spectra of different isotopes or
isotones. Note that the pairing Hamiltonian is also exactly
solvable away from the degenerate limit (i.e., with different
single particle energy levels) but only in the presence of two
orbits \cite{Balantekin:2007ip}. From this solution, we know that
the supersymmetry mentioned above is broken away from the
degenerate limit.

We begin by noting that in the degenerate limit, the one body term
of the pairing Hamiltonian is reduced a constant proportional to
the total number of pairs which can be discarded. Therefore, in
the degenerate limit, the pairing Hamiltonian becomes
\begin{equation}\label{9}
\hat{H}_D=-|G|(\sum_j c^*_{j}S_{j}^{+})%
(\sum_{j^{\prime}}c_{j^{\prime}} S_{j^{\prime}}^-).
\end{equation}
This tells us that the degenerate limit of the pairing Hamiltonian
can also be viewed as the strong pairing limit.

Exact eigenvalues and eigenstates of the Hamiltonian given in Eq.
(\ref{9}) were obtained in Refs.
\cite{Pan:1997rw,Balantekin:2007vs,Balantekin:2007qr} (see Ref.
\cite{Yamac} for the eigenvalues of the quantum invariants of this
Hamiltonian). Following these references, we introduce the
operators
\begin{equation}\label{Define S}
S^+(x)=\sum_j\frac{c^*_j}{1-|c_j|^2 x}S_j^+ \quad\quad
\mbox{and}\quad\quad S^-(x)=\sum_j\frac{c_j}{1-|c_j|^2 x}S_j^-.
\end{equation}
These operators respectively create and annihilate a single pair
of nucleons in the valance shell in a certain distribution over
the single particle energy levels described by the complex
parameter $x$. The eigenstates of the pairing Hamiltonian in the
degenerate limit can be written in terms of these operators. For
example, in Ref. \cite{Talmi} Talmi has showed that, in the
presence of one pair of nucleons in the valance shell,
\begin{equation}\label{A1}
\hat{S}^+(0)|0\rangle \quad\quad\quad E=-|G|\sum_j \Omega_j
|c_j|^2
\end{equation}
is an eigenstate-eigenvalue pair of the Hamiltonian (\ref{9}).
Additional eigenstates with one pair of nucleons can be found by
considering a more general Bethe ansatz state in the form of
$\hat{S}^+(x)|0\rangle$. In fact, using the quasi-spin angular
momentum algebra given in Eq. (\ref{3}), it is easy to show that
\begin{equation}\label{A3}
\hat{S}^+(x)|0\rangle \quad\quad\quad E=0\quad\quad\quad\quad\quad
\end{equation}
is an eigenstate-eigenvalue pair of the Hamiltonian (\ref{9}) if
the parameter $x$ is a solution of the Bethe ansatz equation
\begin{equation}\label{A4}
\sum_j \frac{-\Omega_j/2}{1/|c_j|^2-x}=0.
\end{equation}
Note that Eq. (\ref{A4}) may have several distinct solutions in
which case we have a zero energy eigenstate in the form of Eq.
(\ref{A3}) for each solution. It is worth to emphasize at this
point that all the energies are calculated with respect to the
core and that we have ignored the one body term in the
Hamiltonian.

These considerations generalize to the case of several pairs
occupying the valance shell. Let us denote the maximum number of
pairs which can occupy the valance shell by $N_{max}$. If the
shell is occupied by $N$ pairs such that $N\leq N_{max}/2$ then
the pairing Hamiltonian given in Eq. (\ref{9}) has some
eigenstates with nonzero energy and some eigenstates with zero
energy, similar to the one pair case discussed above
\cite{Pan:1997rw,Balantekin:2007vs,Balantekin:2007qr}. The nonzero
energy eigenstates together with their energies are given by
\begin{equation}\label{A5}
\hat{S}^+(0)\hat{S}^+(z_1) \dots \hat{S}^+(z_{N-1})|0\rangle
\quad\quad E=-|G|\left(\sum_j \Omega_j |c_j|^2-\sum_{k=1}^{N-1}
\frac{2}{z_k}\right)
\end{equation}
where the parameters $z_1,z_2,\dots,z_{N-1}$ are solutions of the
set of Bethe ansatz equations
\begin{equation}\label{A6}
\sum_j \frac{-\Omega_j/2}{1/|c_j|^2-z_m}
=\frac{1}{z_m}+\sum_{k=1(k\neq m)}^{N-1} \frac{1}{z_m-z_k}
\end{equation}
for $m=1,2,\dots,N-1$. Similarly, the zero energy eigenstates are
given by
\begin{equation}\label{A7}
\hat{S}^+(x_1)\hat{S}^+(x_2) \dots \hat{S}^+(x_N)|0\rangle
\quad\quad E=0 \quad\quad\quad\quad\quad\quad\quad\quad
\end{equation}
where the parameters $x_1,x_2,\dots,x_{N}$ are solutions of the
Bethe ansatz equations
\begin{equation}\label{A8}
\sum_j \frac{-\Omega_j/2}{1/|c_j|^2-x_m}=\sum_{k=1(k\neq m)}^N
\frac{1}{x_m-x_k}
\end{equation}
for $m=1,2,\dots,N$. Note that the Bethe ansatz equations given
above for both the non-zero energy eigenstates and the zero energy
eigenstates may have several distinct sets of solutions. In such a
case, each solution set gives us an eigenstate-eigenvalue pair
when substituted in the relevant formulas. In particular, if the
zero energy Bethe ansatz equations have more than one solutions
then the zero energy level is degenerate.

Let us now consider a valance shell which is more than half full,
i.e., $N>N_{max}/2$. One can, in principle, extend the formulas
(\ref{A5}-\ref{A8}) for $N> N_{max}/2$. But in this case one
frequently encounters one of the following two problems: 1) The
Bethe ansatz equations given above yield no solutions or 2) the
Bethe ansatz equations do yield solutions but the corresponding
states vanish identically. For this reason, it is more reasonable
to use hole pairs instead of nucleon pairs to describe a valance
shell which is more than half full.

The pairing Hamiltonian given in Eq. (\ref{9}) is invariant under
a supersymmetry transformation which transforms the spectrum of a
nuclei with $N$ pairs of valance nucleons ($N\leq N_{max}/2$) to
the spectrum of a nuclei with $N-1$ pairs of valance holes. In
order to describe this symmetry, let us first introduce the
particle-hole transformation operator
\begin{equation}
\hat{T}=\exp{\left(-i\pi\sum_j
\frac{\hat{S}_j^++\hat{S}_j^-}{2}\right)}
\end{equation}
$\hat{T}^\dagger$ transforms the empty valance shell $|0\rangle$
into the fully occupied valance shell $|\bar{0}\rangle$ and the
pair creation-annihilation operators into each other, i.e.,
\begin{equation}\label{Transformation}
\hat{T}^\dagger |0\rangle=|\bar{0}\rangle \quad\quad\quad
\hat{T}^\dagger S_j^\pm \hat{T} = S_j^\mp \quad\quad\quad
\hat{T}^\dagger S_j^0 \hat{T} =- S_j^0
\end{equation}
Therefore, it transforms a state containing $N$ particle-pairs
into states containing $N$ hole-pairs. However, there is no
particle-hole symmetry in the problem described by the pairing
Hamiltonian as can be easily verified by showing that
$\left[H_{D},\hat{T}\right]\neq 0$. On the other hand, if we
define the operators
\begin{equation}\label{Susy Generator}
\hat{B}^+=(\sum_{j}c^*_{j} \hat{S}^+_j)\hat{T} \ \ \ \ \ \ \
\hat{B}^-=\hat{T}^\dagger(\sum_{j}c_{j} \hat{S}^-_j),
\end{equation}
then it is easy to show that $\hat{B}^+$ and $\hat{B}^-$ commute
with one another
\begin{equation}
[\hat{B}^+,\hat{B}^-]=0
\end{equation}
and that the pairing Hamiltonian given in Eq. (\ref{9}) can be
written as
\begin{equation}\label{Hamiltonian 7}
\hat{H}_D=-|G|\hat{B}^-\hat{B}^+=-|G|\hat{B}^+\hat{B}^-.
\end{equation}
\begin{wrapfigure}{r}{8cm}
  \hspace*{5mm}\includegraphics[width=7.8cm]{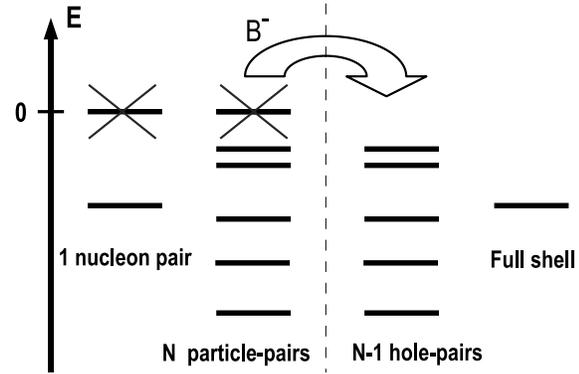}  \\
  \captionsetup{format=plain,labelformat=simple,labelsep=quad,margin=10pt,font=small,labelfont=bf,labelsep=period}
  \caption{Supersymmetry transformation.}\label{Bminus}
\end{wrapfigure}
This tells us that the problem described by the pairing
Hamiltonian in the degenerate limit possesses a quantum mechanical
supersymmetry. Since the operators $\hat{B}^+$ and $\hat{B}^-$
commute with each other, the Hamiltonian $\hat{H}_D$ is equal to
its own supersymmetric partner. As a result, if the state
$|\psi\rangle$ is an eigenstate of the Hamiltonian $\hat{H}_D$
with a nonzero energy, then the state $\hat{B}^-|\psi\rangle$ is
also an eigenstate of $\hat{H}_D$ with the same energy. Zero
energy states, however, are annihilated by $\hat{B}^-$ and do not
transform under the supersymmetry. It is easy to see from the
definition of $\hat{B}^-$ given in Eq. (\ref{Susy Generator}) that
if the state $|\psi\rangle$ has $N$ particle-pairs, then the state
$\hat{B}^-|\psi\rangle$ has $N-1$ hole-pairs. Because the operator
$\hat{B}^-$ first annihilates one particle-pair and then turns the
remaining $N-1$ particle-pairs into hole-pairs. Consider, for
example, the nonzero energy eigenstate with one pair of particles
given in Eq. (\ref{A1}). Using Eqs. (\ref{Transformation}) and
(\ref{Hamiltonian 7}), one can easily show that
\begin{equation}\label{Symmetry N=1}
\hat{B}^-\>\hat{S}^+(0)|0\rangle \propto |\bar{0}\rangle
\end{equation}
Consequently, the states $\hat{S}^+(0)|0\rangle$ and
$|\bar{0}\rangle$ (i.e., the fully occupied valance shell) have
the same energy eigenvalue given by Eq. (\ref{A1}). On the other
hand, the zero energy eigenstates described by Eqs. (\ref{A3}) and
(\ref{A4}) are annihilated by $\hat{B}^-$. Similarly, the nonzero
energy eigenstates with $N$ pairs of nucleons described in Eqs.
(\ref{A5}) and (\ref{A6}) transform as
\begin{equation}
\hat{B}^-\>\hat{S}^+(0)\hat{S}^+(z_1) \dots
\hat{S}^+(z_{N-1})|0\rangle %
\propto \hat{S}^-(z_1) \dots \hat{S}^-(z_{N-1})|0\rangle
\end{equation}
As a result, the states $\hat{S}^+(0)\hat{S}^+(z_1) \dots
\hat{S}^+(z_{N-1})|0\rangle$ and $\hat{S}^-(z_1) \dots
\hat{S}^-(z_{N-1})|0\rangle$ have the same energy given in Eq.
(\ref{A5}). Since $N \leq N_{max}/2$, the former state corresponds
to a nucleus which has less than half full or half full valance
shell whereas the later state corresponds to a nucleus which has
more than half full valance shell (See Figure \ref{Bminus}). Note
that the zero energy states with $N$ pairs of nucleons described
by Eqs. (\ref{A7}) and (\ref{A8}) are annihilated by $\hat{B}^-$.
\begin{wrapfigure}{l}{8cm}\nonumber
\includegraphics[width=7.8cm]{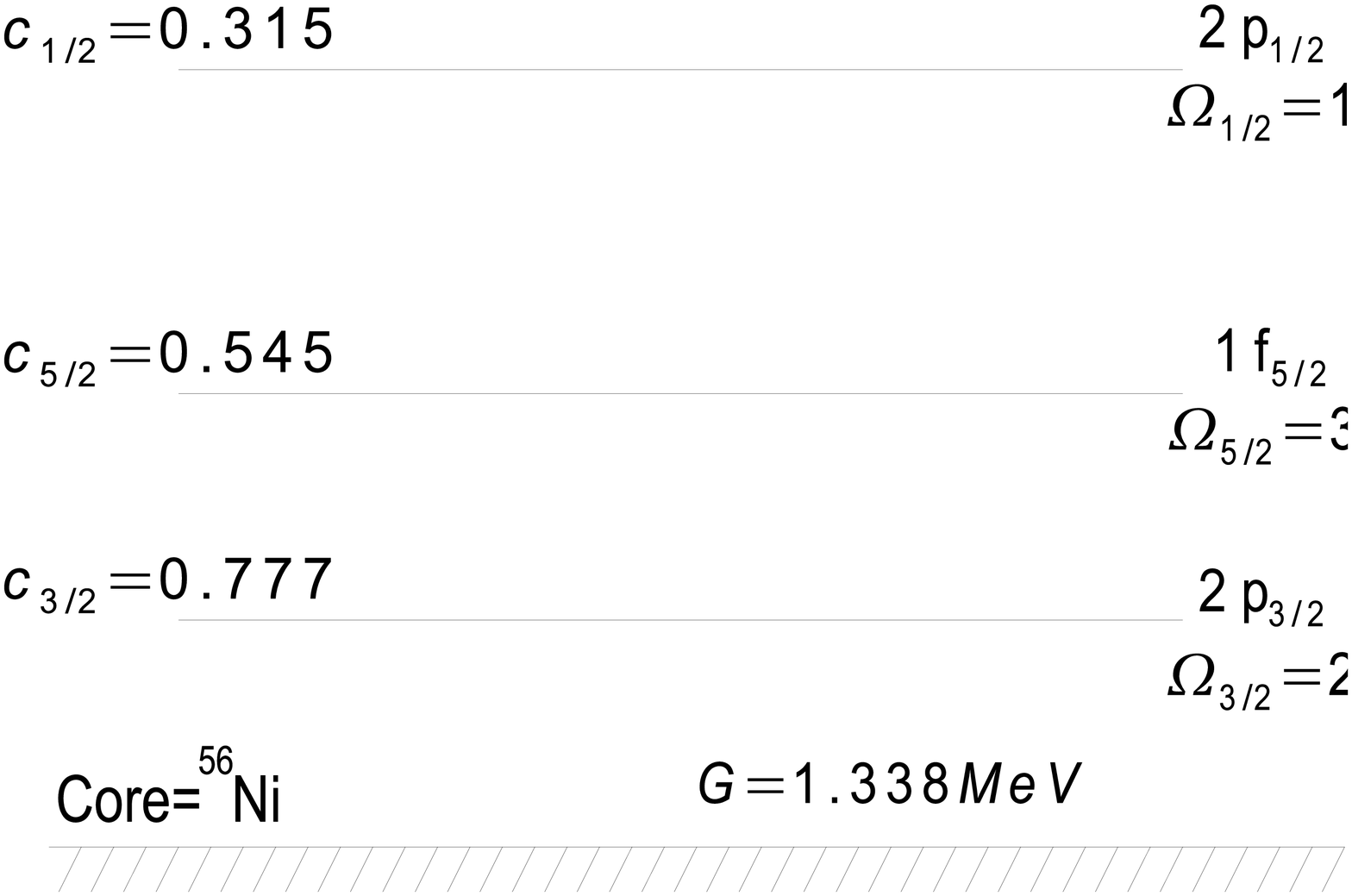}
\captionsetup{format=plain,labelformat=simple,labelsep=quad,margin=10pt,font=small,labelfont=bf,labelsep=period}
\caption{The valance shell for $^{58}$Ni- $^{68}$Ni.}
\label{Nickel}
\end{wrapfigure}

\begin{figure}
\includegraphics[width=12cm, height=7.8cm]{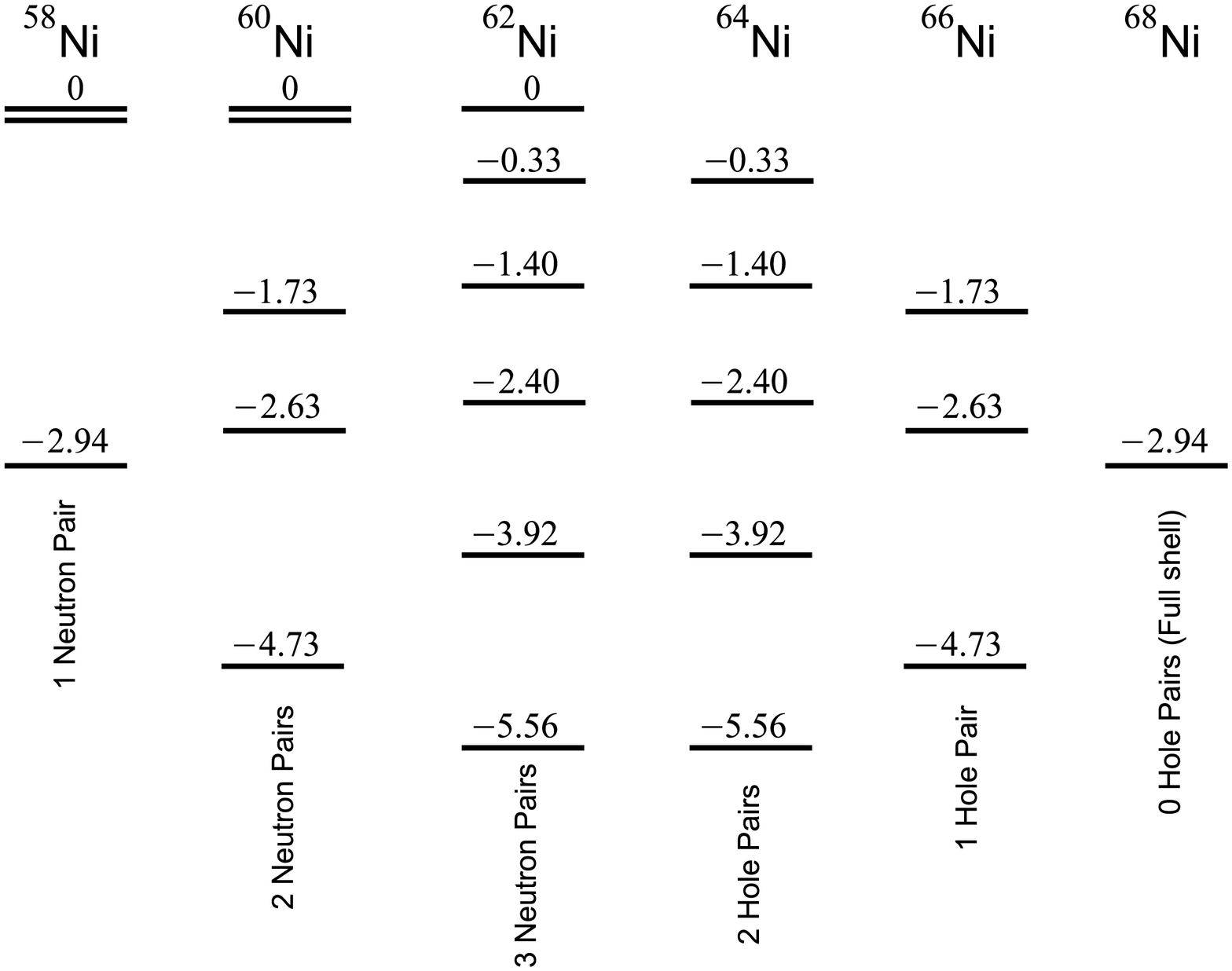}
\captionsetup{format=plain,labelformat=simple,labelsep=quad,margin=10pt,font=small,labelfont=bf,labelsep=period}
\caption{The exact energy eigenvalues of the pairing Hamiltonian
for $^{58}$Ni-$^{68}$Ni in MeV relative to the $^{56}$Ni core. The
degeneracies of the zero energy states of $^{58}$Ni-$^{60}$Ni are
indicated by the double lines. }\label{Ni_Pairing_Energies}
\end{figure}

As an application of this formalism, let us consider the even-even
Ni isotopes from $^{58}$Ni to $^{68}$Ni for which the neutron and
proton shells are closed at $N=Z=28$ (i.e., the core is the doubly
magic $^{56}$Ni nucleus). The extra neutrons in
$^{58}$Ni-$^{68}$Ni occupy the single particle energy levels shown
in Figure \ref{Nickel}. The occupation probability amplitudes
shown in this figure are taken from Ref. \cite{Auerbach} and the
value of the pairing strength $|G|$ is chosen so as to obtain the
ground state energy of $^{58}$Ni (with respect to $^{56}$Ni core)
in agreement with the experimental data. Figure
\ref{Ni_Pairing_Energies} shows the exact energy eigenvalues of
the pairing Hamiltonian given in Eq. (\ref{9}) calculated with the
formalism outlined above. The zero energy states of $^{58}$Ni and
$^{60}$Ni isotopes are doubly degenerate as indicated by double
lines in the figure. All the energies are given in MeV and are
relative to the $^{56}$Ni core. We use the supersymmetry to obtain
the spectra of the nuclei with more than half full shell (i.e.,
$^{64}$Ni-$^{68}$Ni). For example, the spectrum of $^{64}$Ni is
the same as the spectrum of $^{60}$Ni except that the zero energy
states are absent for $^{64}$Ni. See Ref. \cite{Balantekin:2007vs}
for a comparison of these results with the experimentally
available values.

With the Bethe ansatz formalism outlined above, one can easily
calculate the exact numerical values of the pairing energies.
However, to obtain closed analytical expressions for the pairing
energies is relatively difficult in this formalism. Because this
would require the triple effort of solving the Bethe ansatz
equations analytically, substituting the solutions into the energy
expression given in Eq. (\ref{A5}) and then making the necessary
simplifications. In the last part, we present an easier method of
calculating the energy eigenvalues. Although this method is
currently valid for only two orbits, it can possibly be extended
to include more than two orbits.

In the presence of only two orbits, the problem of calculating the
nonzero energy eigenvalues can be greatly simplified by using a
method proposed by Stieltjes \cite{Stieltjes}. Using this method,
one can transform the problem of solving the equations (\ref{A5})
and finding the energy eigenvalues for $N$ nucleon-pairs into the
problem of finding $N^{th}$ order polynomial solutions of the
following differential equation \cite{Nurtac}:
\begin{equation}\label{Heun}
p_{N}^{\prime \prime }\left( z\right) -\left( \frac{\Omega
_{j_2}}{z}+\frac{\Omega _{j_1}}{z-1}\right) p^{\prime }\left(
z\right) +\frac{\left( \alpha \beta z-q_{N}\right) }{z\left(
z-1\right) \left( z-c\right) }p_{N}\left( z\right) =0
\end{equation}
where $\alpha =-N$, $\beta =N-\Omega _{j_1}-\Omega _{j_2}-1$ and
\begin{equation}\label{Parameters}
c=-\frac{1/c_{j_2}^{2}}{1/c_{j_1}^{2}-1/c_{j_2}^{2}} \ \ \ \ \
q=\frac{-E/\left\vert G\right\vert +N\left( N-\Omega _{j_1}-\Omega
_{j_2}-1\right) c_{j_1}^{2}}{c_{j_2}^{2}-c_{j_1}^{2}}.
\end{equation}
Note that a similar but slightly different method was used in Ref.
\cite{Balantekin:2007qr} in order to obtain solutions of the zero
energy Bethe ansatz equations. Eq. (\ref{Heun}) is known as Heun's
differential equation and it admits polynomial solutions only for
certain values of the parameter $q$. One can use the Frobenius
method (i.e., try a series solution) and then find the conditions
on $q$ such that the series will terminate at the $N^{th}$ order.
This gives us the allowed values of the energy eigenvalues for $N$
pairs of nucleons because $q$ and $E$ are related by Eq.
(\ref{Parameters}). This method significantly reduces the amount
of calculation needed to compute the energy eigenvalues of the
pairing Hamiltonian given in Eq. (\ref{9}) \cite{Nurtac}.

\noindent This  work  was supported in  part  by   the  U.S.
National Science Foundation Grant No. PHY-0555231 at the
University of  Wisconsin, and  in  part by  the University of
Wisconsin Research Committee   with  funds  granted by the
Wisconsin Alumni  Research Foundation.


\begin{thebibliography}{99}

\bibitem{Kerman1}
A.~K.~Kerman, Ann. Phys. {\bf 12} (1961), 300.

\bibitem{Balantekin:2007ip}
  A.~B.~Balantekin and Y.~Pehlivan,
  Phys.\ Rev.\  C {\bf 76}, 051001 (2007)
  [arXiv:0710.3941 [nucl-th]].

\bibitem{Pan:1997rw}
  F.~Pan, J.~P.~Draayer and W.~E.~Ormand,
  Phys.\ Lett.\  B {\bf 422}, 1 (1998)
  [arXiv:nucl-th/9709036].

\bibitem{Balantekin:2007vs}
  A.~B.~Balantekin, J.~H.~de Jesus and Y.~Pehlivan,
  Phys.\ Rev.\  C {\bf 75}, 064304 (2007)
  [arXiv:nucl-th/0702059].

\bibitem{Balantekin:2007qr}
  A.~B.~Balantekin and Y.~Pehlivan,
  J.\ Phys.\ G {\bf 34}, 1783 (2007)
  [arXiv:0705.1318 [nucl-th]].

\bibitem{Yamac}
  Y.~Pehlivan, [arXiv:0806.1810 [math-ph]].

\bibitem{Talmi} I. Talmi, Nucl. Phys. {\bf A172}, 1 (1971)

\bibitem{Auerbach} N. Auerbach, Nuclear Physics {\bf 76} (1966) 321-335.

\bibitem{Stieltjes} T.J. Stieltjes, 1914, {\em Sur Quelques Theoremess
d'Algebre, Oeuvres Completes}, V. 11 (Groningen:Noordhoff).

\bibitem{Nurtac} N.~G{\"{u}}ven and Y.~Pehlivan., in preparation.

\bibitem{heun}  A. Ronveaux, 1995, {\em Heun's Differential
Equations}, (Oxford Science Publications).

\end{thebibliography}
\end{document}